\documentclass[letterpaper,11pt,oneside,reqno]{amsart}

\usepackage{bm}
\usepackage{cite}
\usepackage{caption}
\usepackage{afterpage}
\usepackage{enumitem}
\usepackage{bmpsize}
\usepackage{hyperref}
\usepackage{subcaption}

\usepackage{amsmath,amssymb,amsthm,amsfonts,dsfont}
\usepackage{hyperref}
\usepackage{graphicx,color}
\usepackage{upgreek}
\usepackage[mathscr]{euscript}

\allowdisplaybreaks
\numberwithin{equation}{section}

\usepackage{tikz}
\usetikzlibrary{shapes,arrows,positioning,decorations.markings}

\usepackage{float}
\usepackage{multicol}
\usepackage{array}
\usepackage{adjustbox}
\usepackage{cleveref}
\usepackage{enumerate}

\usepackage[DIV12]{typearea}

\newcommand{\etalchar}[1]{$^{#1}$}

\usepackage{comment}

\theoremstyle{plain}

\theoremstyle{plain}
\newtheorem{theorem}{Theorem}[section]

\theoremstyle{definition}

\newtheorem{rem}[theorem]{Remark}

	\tikzstyle{vecArrow} = [thick, decoration={markings,mark=at position
   1 with {\arrow[semithick]{open triangle 60}}},
   double distance=1.4pt, shorten >= 5.5pt,
   preaction = {decorate},
   postaction = {draw,line width=1.4pt, white,shorten >= 4.5pt}]

\begin{document}
\title{The KPZ Universality Class and Related Topics}

\author[A. Saenz]{Axel Saenz}
\address{A. Saenz, University of Virginia, Department of Mathematics,
141 Cabell Drive, Kerchof Hall,
P.O. Box 400137,
Charlottesville, VA 22904, USA
}
\email{ais6a@virginia.edu}

\date{}

\maketitle

\begin{abstract}
	These notes are based on a talk given at the \emph{2018 Arizona School of Analysis and Mathematical Physics}. We give a comprehensive introduction to the \emph{KPZ universality class}, a conjectured class of stochastic process with local interactions related to random growth processes in $1+1$ dimensions. We describe some of the characteristic properties of the KPZ universality class such as scaling exponents and limiting statistics. In particular, we aim to extract the characteristic properties of the KPZ universality class by understanding the KPZ stochastic partial differential equation by a special discrete approximation given by the asymmetric simple exclusion process (ASEP). The connection with the ASEP is very important as the process enjoys a rich integrability structure that leads to many exact formulas.
\end{abstract}

\setcounter{tocdepth}{3}
\tableofcontents
\setcounter{tocdepth}{3}

\section{Introduction}
\label{sec:intro}

	The \textit{KPZ universality class} was born on 1986 in the work of physicists  M. Kardar, G. Parisi, and Y.C. Zhang in \cite{kpz}, where the authors `` ... described how a growing interface can be studied by dynamic renormalization." The growing interface is described by a height function $h(t,x)$ with spatial variable $x \in \mathbb{R}^n$ for some $n \in \mathbb{N}$ and a time variable $t \in \mathbb{R}_{\geq 0}$. The height function for $n=1$ will be a focal element of the KPZ universality class. In the work of M. Kardar, G. Parisi and Y.C. Zhang, the scaling exponents of the height function $h(t,x)$ are derived depending on the dimension variable $n$. The authors assumed that the evolution of the height function $h(t,x)$ is a solution to the \textit{stochastic partial differential equation (SPDE)}: 
\begin{equation}\label{kpz}
\partial_t h(t,x) = \nu \partial^2_x h(t,x) + \frac{\lambda}{2} \left( \partial_x h(t,x) \right)^2 + W(t,x),
\end{equation}
with $W(t,x)$ a Gaussian space-time white noise random variable (i.e.~a $\delta$-correlated stationary, Gaussian process with mean zero and covariance $\mathbb{E}\left[ W(t' , x') W(t,x) \right] = \delta(t'-t) \delta(x'-x)$). The first term $\nu \partial^2_x h(t,x)$ is a diffusion term which tends to smooth out the height function as time evolves, and the term $\frac{\lambda}{2} \left( \partial_x h(t,x) \right)^2$ is a lateral growth term, which makes the surface grow normal to its interface. The SPDE defined by eqn.~\eqref{kpz} is now called the \emph{KPZ equation}.\par

	It turns out that solving the KPZ equation is highly non-trivial, starting with making the KPZ equation \eqref{kpz} well-defined. For instance, through a local scaling argument, one might expect that the height function behaves similar to a \emph{Brownian motion} due to the space-time white noise, and this means that, a priori, the term $\partial_x h(t,x)$ is a distribution without a well-defined square $(\partial_x h(t,x))^2$ making the KPZ equation ill-defined. Under certain initial conditions, this issue has been resolved in the work of \cite{BerGia, ACQ} by introducing the Cole-Hopf transformation (see eqn.~\ref{CH}) and an appropriate discrete approximation of the KPZ equation, and also, in the work of \cite{HairerKPZ} through the theory of rough paths. In \cite{BerGia}, Bertini and Giacomoni gave a solution to the KPZ equation for initial conditions $h(0, x)$ that (in average) grow at most linearly, such as flat initial conditions or two-sided Brownian motion initial conditions. In particular, Bertini and Giacomoni approach the KPZ equation by first smoothing out the noise, solving the smoothed KPZ equation via a discrete approximation using a discrete Cole-Hopf transformation, and finally taking the appropriate limits to a solution of the KPZ equation. Then, in \cite{ACQ}, Amir, Corwin and Quaste were able to extend the results of \cite{BerGia} by treating the KPZ equation with the infinite narrow wedge initial condition (i.e.~$h(0,x) = \log (\delta_{x=0})$), which is a natural initial condition under the random polymer interpretation of the KPZ equation. In \cite{ACQ}, the authors still use the idea of considering the Cole-Hopf transformation and were able to treat the wedge initial conditions via the exact formulas, introduced by Tracy and Widom in \cite{TW, TW09}, for the \emph{asymmetric simple exclusion process (ASEP)}, which serves as the proper discretization for the KPZ equation. Now, most recently in \cite{HairerKPZ}, Hairer gave an ``path-wise" solution of the KPZ equation for $\alpha$-H\"older initial conditions with $\alpha \in (0, 1/2)$ that is better suited for approximation among different solutions. Moreover, Hairer's solution is independent of the Cole-Hopf transformation and the ASEP approximation of the KPZ equation and also shows that the rough path solution agrees with the Cole-Hopf solution under $\alpha$-H\"older initial conditions. Lastly, we must note that solving the KPZ equation under general initial conditions is still an open problem and an active area of research. In this review, we will focus on the approximation approach to the KPZ equation, similar to that in \cite{BerGia, ACQ}, for the wedge initial condition as it requires less background and it helps build concrete intuition for the KPZ equation.\par

	Our main objective is to understand the probability distribution of the height function $h(t,x)$ for $x \in \mathbb{R}$ as $t \rightarrow \infty$ (by some type of \textit{central limit theorems} such as \emph{limit shapes}). Take the KPZ 1:2:3-scaling:
	\begin{equation}\label{eq:scalingKPZ}
	\epsilon^{1/2} h(\epsilon^{-3/2} t, \epsilon^{-1} x) - C_{\epsilon} t,
	\end{equation}
for $\epsilon \rightarrow 0^{+}$ and some constant $C_{\epsilon} \in \mathbb{R}$ depending on $\epsilon$ (such as $C_{\epsilon} = \epsilon^{-1}/2$). Then, the height function $h^{TASEP}(x, t)$ for the TASEP (see \ref{sec:asep}) under the KPZ 1:2:3-scaling \eqref{eq:scalingKPZ} has a limit to an invariant limiting process $\mathfrak{h}(t,x)$ called the \emph{KPZ fixed point}, recently defined by Matetski, Quastel and Remenik in \cite{matetski2016}. In fact, the \emph{strong KPZ universality conjecture} states that ``the [KPZ fixed point $\mathfrak{h}$] is the limit under the [KPZ 1:2:3-scaling~\eqref{eq:scalingKPZ}] for any model in the [KPZ universality class], loosely characterized by having: 1.~Local dynamics; 2.~Soothing mechanism; 3.~Slope dependent growth rate (lateral growth); 4.~Space-time random forcing with rapid decay of correlations..." as given in \cite{matetski2016}. The strong KPZ universality conjecture is still wide open.\par
   
\begin{figure}[h]
\centering
\includegraphics[width=1\textwidth]{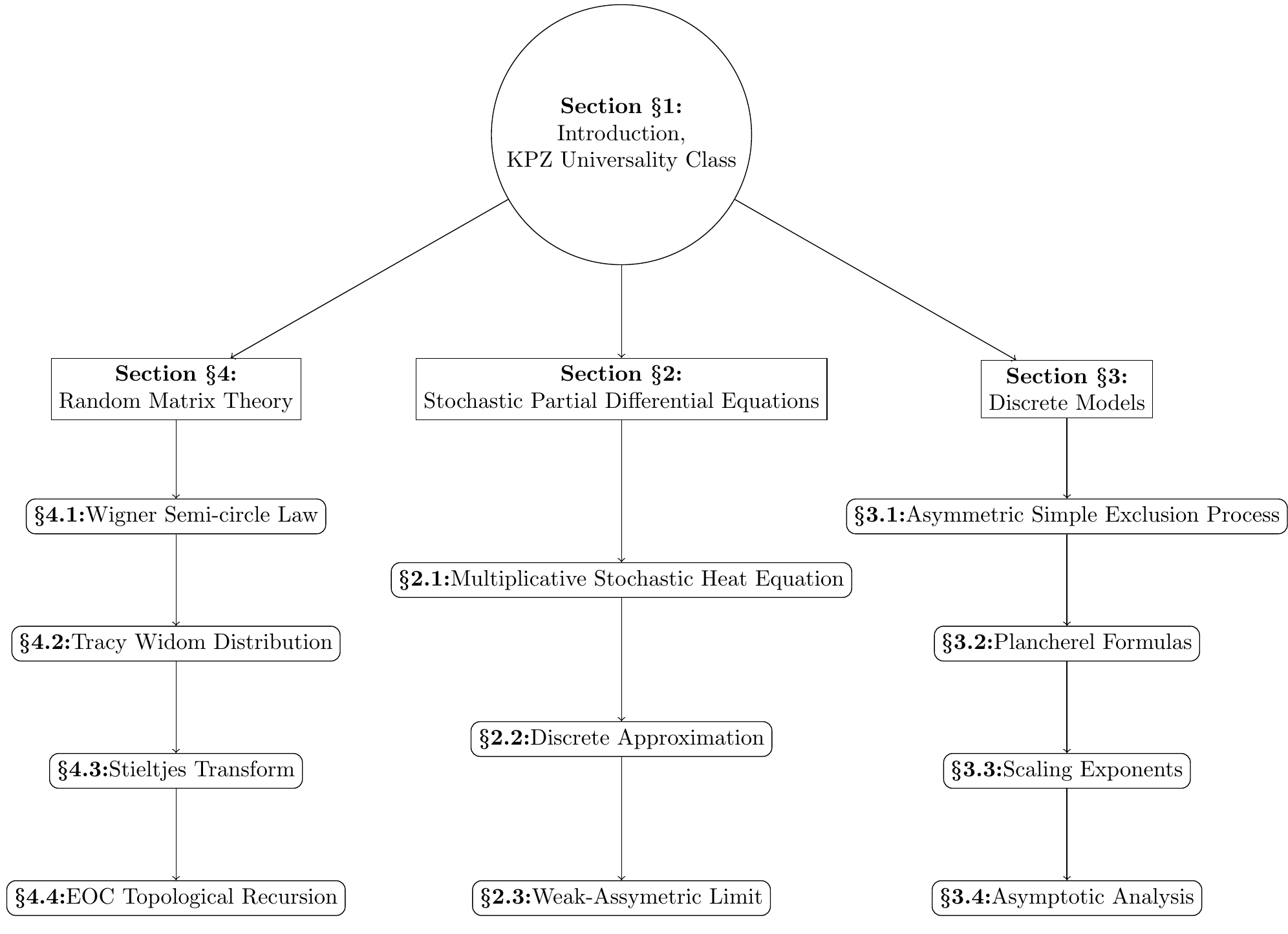}
\caption{A rough outline of the content.}
\label{fig:outline}
\end{figure} 

	In the following sections, our goal is to give enough background to understand the different components of the strong KPZ universality conjecture. In the process, we will point out certain characteristics of~\eqref{kpz} that have appeared in other models that also have similar central limit theorems, giving a non-rigorous but practical description of \textit{KPZ universality class}. We will focus on three subjects: \emph{discrete models}, \emph{stochastic partial differential equations}, and \emph{random matrix theory}. Discrete models include interacting particles systems such as the ASEP. These models have discrete height functions (e.g.~$h^{ASEP}(t,x)$ for ASEP) that approximate the height function for KPZ $h(t,x)$ in some sense (see \ref{sec:Bertini}). The limit of the ASEP to the KPZ equation is described along with the discussion of the SPDEs. In the Random Matrix Theory discussion, we discuss some classical results on limit shapes such as the \emph{Wigner semi-circle law}, and in this same setting, we are able to describe the Tracy-Widom distribution as it was originally discovered by Tracy-Widom in \cite{TW94}. We also discuss other topics related asymptotic analysis of random matrices. The rough outline of the paper if given by Fig.~\ref{fig:outline}.

\section{Stochastic Partial Differential Equations}
\label{sec:SPDEs}

\subsection{Multiplicative Stochastic Heat Equation}
\label{sec:MSHE}

	A priori, the KPZ equation~\eqref{kpz} is not well-defined since the product $(\partial_x h(t, x))^2$ may not be well-defined. The random forcing term $W(t,x)$ may cause a (possible) solution $h(t,x)$ of eqn.~\eqref{kpz} to be a distribution in a H\"older space where the product $(\partial_x h(t,x))^2$ is not defined. Indeed, one must define the product $(\partial_x h(t,x))^2$ by the methods of regularity structures or rough paths developed by Hairer and Lyons (see \cite{Hairer14,Lyons98}) to make the KPZ equation well-defined. In this review, we avoid the theory of regularity structures and rough paths in order to keep the discussion at a more elementary level, and instead, we assume the solution of eqn.~\eqref{kpz} to be given as a limit of smooth functions corresponding to solutions of eqn.~\eqref{kpz} with smooth mollified noise. This approach is much closer to the work of Bertini and Giacomin in \cite{BerGia}. We take the \emph{stochastic heat equation (SHE) with multiplicative white noise},
	\begin{equation}\label{heat}
	\partial_t Z(t, x) = \frac{1}{2} \partial_x^2 Z(t, x) + Z(t, x) W(t,x),
	\end{equation}
as our starting point. For now at a formal level, we have that the KPZ equation is related to the multiplicative SHE via the \emph{Cole-Hopf transformation},
	\begin{equation}\label{CH}
	h(t,x)= - \log Z(t,x).
	\end{equation}
We will have that under certain conditions (i.e.~if $Z(t,x) >0$) the solutions to the stochastic heat equation will make the Cole-Hopf transformation legitimate. In the guise of the multiplicative SHE, we study the KPZ equation using ideas and formulas inspired by similar results in the theory of parabolic \emph{partial differential equations} for the stochastic heat equation (see \cite{Krylov96}).\par

	 Let's recall some results for the \emph{homogeneous heat equation}, 
	\begin{equation}\label{eq:heat}
	\partial_t Z = \frac{1}{2} \partial_x^2 Z.
	\end{equation}
One may solve eqn.~\eqref{eq:heat} by a convolution formula. Define the function
	\begin{equation}\label{eq:fundamental}
	J(t,x) := \int_{\mathbb{R}} \rho(t, x-y) Z(t,y) dy
	\end{equation}
with $\rho(t,x) = \frac{e^{- x^2 /2t}}{\sqrt{2 \pi t}}$. Then, $J(t,x)$ is a solution to the heat equation~\eqref{eq:heat} with the initial conditions $Z(0,x)$. We call $J(t,x)$ the \emph{fundamental solution} given by eqn.~\eqref{eq:fundamental}. This solution is obtained by taking the Fourier transform of eqn.~\eqref{eq:heat}, solving the dual equation, and taking the inverse Fourier transform (see \cite{Krylov96}). In general, the solution of the \emph{inhomogeneous} heat equation (i.e.~with a forcing term $f(t,x)$),
	\begin{equation}
	\partial_t Z = \frac{1}{2} \partial_x^2 Z + f(t,x),
	\end{equation}
may be written similarly as
	\begin{equation}
	Z(t,x) = J(t, x) + \int_{s}^t \int_{\mathbb{R}} \rho(t-u, x-y)f(u,y)du dy.
	\end{equation}
We use similar formulas to analyze the stochastic heat equation with multiplicative white noise, and the computations may be rigorously proven with technical conditions such as $\lim_{t\rightarrow 0^+} \rho(t,x) = \delta_{x}(0)$ in distribution (fast enough) (see \cite{nualart2014}). Using the \emph{Stratonovich integral}, we take 
	\begin{equation}\label{intHeat}
	Z(t,x) = J(t, x) +  \int_{s}^t \int_{\mathbb{R}} \rho(t-u, x-y)Z(u,y)W(u, y)
	\end{equation}
for all $0 \leq s <t$, and we require  $Z(t,x)$ is $\sigma(\{ W(s,\cdot) | 0 \leq s \leq t\} )$-measurable. The function~\eqref{intHeat} is called a \textit{mild solutions} of eqn.~\eqref{heat} (see \cite{nualart2014}). This will be the solution we will consider henceforth.

\begin{theorem}[C. Mueller \cite{Mueller}] Let $Z(t,x)$ be a solution to~\eqref{heat} with a $Z(0, x) \geq 0$ and $\int_{\mathbb{R}} Z(t,0) >0$. Then, for all $t>0$, $Z(t,x) > 0$ for all $x \in \mathbb{R}$ with probability one.
\end{theorem}

The proof follows form using the form of the mild solution~\eqref{intHeat} and large deviation estimates (a sketch can be found in \cite{Quastel}). More importantly, this theorem now legitimizes the Cole-Hopf transformation~\eqref{CH} as we are now taking the $\log$ of a strictly positive number. Therefore, one may wish to think of the KPZ equation as the stochastic heat equation with multiplicative white noise.

\begin{rem}
	One may obtain the \emph{Wiener chaos expansion} of the mild solution to the multiplicative SHE~\eqref{heat} by iterating the solution given in eqn.~\eqref{intHeat}:
	\begin{equation}\label{eq:wiener}
	\begin{split}
	&Z(t_0,x_0) = J(t_0, x_0) +\\
	& \hspace{-5mm} \sum_{n \geq 1} \int_0^{t_0} \cdots \int_0^{t_{n-1}} \int_{\mathbb{R}^n} \rho(t_0 -t_1, x_0-x_1) \cdots \rho(t_{n-1} - t_n , x_{n-1} -x_n) J_{0}(t_n, x_n) W(t_n ,  x_n) \cdots W( t_1, x_1).
	\end{split}
	\end{equation}
See \cite{lechen} for a detailed definition of the Wiener chaos expansion.~\eqref{eq:wiener} along with some specific properties and detailed conditions on the initial conditions $Z(0,x)$.
\end{rem}

\begin{rem}
	In the theory of directed polymers, one can show that the partition function of the model is a solution to the stochastic heat equation. In a discrete way, these polymers may be thought of up-right paths on the positive integer lattice $\mathbb{Z}_+ \times \mathbb{Z}_+$. Each lattice point has weight given by random variable where all weights are independent and identically distributed with exponential distribution. Then, each path (each being equally likely) has an energy given by the sum of the weights of the lattice points along its path. The partition function is then given as a sum of the Boltzmann weights of all the possible paths, and one may also considers a restriction that the path be taken from one geometry (such as a point or a line) to another geometry (such as another point or another line). This discrete model has been shown to be in the KPZ universality class and by taking the continuum limit \cite{ACQ}, one has that the partition function is a solution of the stochastic heat equation with multiplicative noise. There are many exact result for these models under an \emph{integrable} setting, and the interested reader should consult \cite{BorCorMac} for a brief introduction to directed polymers along with their moment formulas. We just like to highlight that the stochastic heat equation has an interpretation of a random growth model on its own right and such model also enjoy the properties of the KPZ universality class.
\end{rem}

\subsection{Discrete Approximation}
\label{sec:Bertini}

	In section~\ref{sec:MSHE}, the Cole-Hopf transformation~\eqref{CH} defines a solution of the KPZ equation inspired by well-known ideas and formulas for SPDEs. This lets us know that the height function of the KPZ equation does indeed exist after a proper interpretation via the Cole-Hopf transformation. Now, we construct a solution of the KPZ equation~\eqref{kpz}. Instead of solving the KPZ equation via the Cole-Hopf transformation, we consider a solution to the KPZ given by a scaling limit of a discrete approximation, first given in the work of Betrtin and Giacomin \cite{BerGia} and further enhanced by the work of Amir, Corwin and Quastel \cite{ACQ} using the hallmark results of Tracy and Widom \cite{TW, TW09}.\par

At its most basic level, in the scaling limit approach, one takes a discrete approximation of the KPZ equation, solves the discrete problem, takes a scaling limit, and checks the limiting object to be a solution of the KPZ equation. Of course, there are many subtleties in carrying out all of aforementioned steps correctly, and one should see \cite{ACQ, dembo16, corwin18} for a successful attempt. It turns out, there are many discrete models that have limiting KPZ statistics and seem to be in the KPZ universality class (e.g.~last passage percolation, polyneuclar growth, six vertex model, polymers in random medium, ASEP). The discrete models tend to behave similarly (which can be seen via the KPZ scaling~\eqref{eq:scalingKPZ} of their perspective one point function \cite{Corwin}) with bijections through explicit of transformations or equivalence of moment formulas \cite{BorCor}.\par

	In this section, we will sketch an argument showing that the ASEP model (see section~\ref{sec:asep} for a basic introduction) is a discretization of the KPZ equation~\eqref{kpz} under certain conditions. The rigorous treatment of the previous statement is highly technical and the full details may be found in the work of Bertini Giacomin \cite{BerGia}. Thus, we leave many of the technical details out and refer the reader to \cite{BerGia} for the precise statements and details. We show that the evolution formulas for ASEP resemble the evolution formulas for KPZ before we take the scaling limit. In order to actually take the scaling limit, one must prove some specific inequalities for the limits to behave well.\par

	We start with the height function $h^{ASEP}(t,x)$ for the ASEP, and show that is satisfies a discrete version of the KPZ equation~\eqref{kpz}. Let $\vec{x}(t) = \{ x_1(t), \dots, x_N(t) \}$ be a particle configuration of the ASEP on the integer lattice $\mathbb{Z}+\frac{1}{2}$ at time $t$, and introduce the \emph{occupation variable}
	\begin{equation}\label{eq:occupation}
	\eta_t (x) := \mathds{1}(x \in \vec{x}(t))
	\end{equation}
to be the indicator function that there is some particle at location $x \in \mathbb{Z} + \frac{1}{2}$ in the configuration $\vec{x}(t)$. Then, we set the height function at zero to be given by twice the number of particles to the right of the origin,
	\begin{equation}\label{eq:ASEP_height_zero}
	h^{ASEP}(t,0) := 2 \sum_{y = 1/2}^{\infty} \eta_t(y),
	\end{equation}
and in general for any $x \in \mathbb{Z}$, the height function is given by
	\begin{equation}\label{eq:ASEP_height}
	h^{ASEP}(t,x) :=  \begin{cases} h^{ASEP}(t,0)  +\overset{x - 1/2}{\underset{y=1/2}{\sum}}\bigg(  1 - 2 \eta_t(y) \bigg), \quad \hspace{2mm} x \geq 1 \\  h^{ASEP}(t,0)  - \overset{x + 1/2}{\underset{y=-1/2}{\sum}} \bigg( 1 - 2 \eta_t(y) \bigg) , \quad x \leq -1\end{cases}.
	\end{equation}
Additionally, we extend $h^{ASEP}(t, x)$ to all $x \in \mathbb{R}$ by linear interpolation among the values for $x \in \mathbb{Z}$. The notation $h^{ASEP}(t,x)$ for the height function of ASEP is used to distinguish it from the height function $h(t,x)$ for the KPZ equation~\eqref{kpz}.\par

\begin{figure}[h]
\begin{subfigure}{0.45\textwidth}
\includegraphics[width =\textwidth]{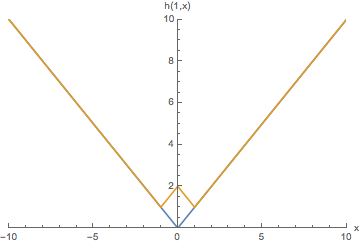}
\caption{A simulation of the height function for the ASEP model at $t=1$.}
\end{subfigure}
\begin{subfigure}{0.45\textwidth}
\includegraphics[width=\textwidth]{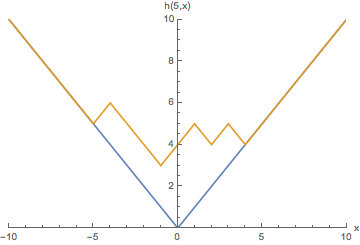}
\caption{A simulation of the height function of the ASEP model at $t=5$.}
\end{subfigure}
\caption{}
\end{figure}

	Now, take the height function $h(t, x)$ of the KPZ equation~\eqref{kpz} and (formally) set $u(t, x) = \partial_x h(t,x)$. If $h(t, x)$ satisfies the  KPZ equation~\eqref{kpz}, then transformation $u(t,x)$ satisfies the \emph{stochastic Burgers equation},
	\begin{equation}\label{burgers}
	\partial_t  u(t,x) = \nu \partial_x^2 u(t,x) + \lambda u(t,x) \partial_x u(t,x) + \partial_x W(t,x).
	\end{equation}
We don't worrying too much about the definition $\partial_x W(t,x)$ since we won't actually take the scaling limit, and we simply identify that the classical Burgers' equation has a random forcing term $ \partial_x W(t,x)$. Then, note that the discrete derivative of the ASEP height function $h^{ASEP}(t,x)$ is given by $h^{ASEP}(t, n+1) - h^{ASEP}(t, n)  =  1 - 2\eta_t(n)$. Thus, we want to show that 
	\begin{equation} \label{eq:disc_der}
	u(t,n):=1 - 2 \eta_t(n)
	\end{equation}
satisfies a discrete Burgers' equation
	\begin{equation}\label{discrete}
	\partial_t u(t, n)  = \nu \big[u(t, n+1) - 2 u(t,n), +u(t,n) \big] + \lambda u(t,n) \frac{ u(t, n+1) - u(t,n-1) }{2} + X(t,n),
	\end{equation}
with $X(t,n)$ a $\sigma \left( X(t-1, \cdot)\right)$-measurable random variable that converges to $\partial_x W(t,x)$ under the proper scaling. In particular, with the appropriate scaling and certain conditions on the regularity of $u(t,n)$, the first term on the right side of eqn.~\eqref{discrete} will converge to the second partial derivative with respect to space, the second term on the right side eqn.~\eqref{discrete} will converge to the product of the function and the partial derivative with respect to space, and the last term will converge to the random forcing term. Therefore, we take equation~\eqref{discrete} as the discrete version of the Burgers' equation~\eqref{burgers}. Moreover, by appealing to local interaction of particles in ASEP for small time increments $\Delta t$, we derive a differential-difference equation for the occupation variable $\eta_x$ that is equivalent to the Kolmorov forward master equation written in coordinate form when we take $\Delta t \rightarrow 0$.\par
	Consider the time evolution of the the occupation variable $\eta_t (x) := \mathds{1}(x \in \vec{x}(t))$ for the ASEP. The particles in the ASEP evolve according to a sequence of independent exponential clocks $T_x$ (with rate $1$) for $x \in \mathbb{Z} + \frac{1}{2}$. Each of the exponential clocks is actually defined by two other independent exponential clocks $T_{x,+}$ and $T_{x, -}$ (i.e.~$T_x = \text{min}(T_{x,-}, T_{x,+})$) with rates $p$ and $q$, respectively, such that $p+q = 1$. If there is a particle on site $x$, the particle will attempt to move to an adjacent location on $\mathbb{Z} + \frac{1}{2}$ depending $T_{x,+} = T_x $ or $T_{x,-} = T_x$. For $T_{x,\pm} = T_x $ and a particle at site $x$ (if there is a particle) will move to site $x\pm1$ if the location is available. In particular, we consider three cases for the evolution of $u(t,n)$to $u(t + \Delta , n)$ and $n \in \mathbb{Z} + 1/2$: 
	\begin{itemize}
	\item[(i)] there is no particle at site $n$ allowing particles from the left and right to jump in; 
	\item[(ii)] there is a particle in site $n$ and it may or may not be able to jump to a neighboring site;
	\item[(iii)] or the clock doesn't go off. 
	\end{itemize}
In the following, we only consider the interaction of the locations $x-1, x, x+1$.\par

	For case (i), there is no particle at location $x$ meaning that the occupation variable $\eta_t(x)$ will only evolve by interaction with particles at locations $x\pm1$. For small enough $\Delta t$, we have 
	\begin{equation}
	\begin{split}
	&\eta_{t+ \Delta t}(x)= \\
	& \eta_{t}(x+1) \mathds{1}(T_{x+1, -} = T_{x+1, -} \wedge T_{x-1, +} < \Delta t) + \eta_t( x-1) \mathds{1}(T_{x-1, +} = T_{x+1, -} \wedge T_{x-1, +} <\Delta t)  + o(\Delta t),
	\end{split}
	\end{equation}
with the first term on the right side denoting a particle moving from $x+1$ to $x$ and the second term denoting a particle moving from $x-1$ to $x$. If there is no particle, say, on the site $x-1$, then $\eta_t( x-1) =0$ causing no change even if the exponential clock for the particle at site $x-1$ is activated, and similarly for a particle on site $x+1$.\par

	For case (ii), there is a particle on site $x$, we have
	\begin{equation}
	\eta_{t+ \Delta t}( x) = \eta_t(x+1) \mathds{1}(T_{x, +} = T_{x} <\Delta t) + \eta_t(x-1) \mathds{1}(T_{x, -} = T_{x} < \Delta t) + o(\Delta t),
	\end{equation}
with the first term on the right side denoting a particle moving from $x$ to $x+1$ and the second term denoting a jump from $x$ to $x-1$. Since there is a particle at site $x$, the occupation variable $\eta_t(x)$ will only change if the particle at site $x$ moves regardless of the activation of the particles at site $x-1$ or $x+1$.\par

	For case (ii), nothing happens, we simply have that
	\begin{equation}
	\eta_{t+ \Delta t}( x) = \eta_t( x) \mathds{1}( T_{x} \geq \Delta t) = \eta_t( x)  -\eta_t( x) \mathds{1}( T_{x} < \Delta t).
	\end{equation}
Note that case (ii) only happens when $\eta_t( x)= 1$, and case (i) happens when $\eta(t,n) =0$. We combine all cases into a single equation:
	\begin{equation}\label{eq:evol}
	\begin{split}
	&\eta_{t+ \Delta t} (x) - \eta_{t}(x)  \\
	&\hspace{-7mm}=  \big[\eta_{t}(x+1) \mathds{1}(T_{x+1, -} = T_{x+1, -} \wedge T_{x-1, +} <\Delta t) + \eta_t( x-1) \mathds{1}(T_{x-1, +} = T_{x+1, -} \wedge T_{x-1, +} <\Delta t)\big](1- \eta_t(x) )\\
& + \big[ \eta_t(x+1) \mathds{1}(T_{x, +} = T_{x} < \Delta t) + \eta_t( x-1) \mathds{1}(T_{x, -} = T_{x} < \Delta t) \big] \eta_t(x) - \eta_t(x) \mathds{1}( T_{x} < \Delta t) .
	\end{split}
	\end{equation}
Lastly, we will center the random variables $\mathds{1}(T_x,\pm = T > \Delta t)$ by subtracting their mean. That is,
	\begin{equation}
\mathds{1}(T_{x,+} = T > \Delta t) = p \Delta t+ Y_x , \hspace{5mm} \mathds{1}(T_{x,-} = T > \Delta t) = q \Delta t + \tilde{Y}_x.
	\end{equation}
Then, we take the $\Delta t \rightarrow 0$ limit of equation \eqref{eq:evol} and we obtain
	\begin{equation}
	\begin{split}
	\partial_t \eta_{t}(x) &= q \eta_t(x+1) - \eta_t(x) + p \eta_t(x-1)\\
	&= (p-q) \eta_t(x) \bigg[ \eta_t(x+1) - \eta_t(x-1)  \bigg] + Z(x)
	\end{split}
	\end{equation}
by accumulating all of the random variables into the single random variable $Z(x)$ that is $\eta_t$-measurable. (It is left as an exercise to the reader to write $Z(x)$ in terms of the $\eta_t$-measurable random variables $T_{x,\pm}$'s and $\eta_t(x)$'s.) We should have that $Z(x) \rightarrow \partial_x W(t,x)$ under an appropriate scaling limit. In particular, for small $\epsilon >0$, we set the \emph{weak asymmetric limit} $p-q = \lambda \epsilon^{1/2}$ and define 
	\begin{equation}\label{eq:discrete2}
	u^{\epsilon} (T, X) := \epsilon^{-1/2} \bigg(1- 2 \eta_{ \epsilon^{-2}T}(\lfloor\epsilon^{-1} X \rfloor) \bigg)
	\end{equation}
for $X \in \mathbb{R}$ and $T \in \mathbb{R}_{\geq 0}$. Then, we have
	\begin{equation}\label{eq:burgers2}
	\partial_T u^{\epsilon} (T, X) = - \lambda u^{\epsilon}(T, X) \frac{u^{\epsilon}(X +\epsilon)  - u^{\epsilon}(X +\epsilon) }{2 \epsilon} + \frac{u^{\epsilon}(X +\epsilon) - 2 u^{\epsilon}(X ) +u^{\epsilon}(X -\epsilon)}{2 \epsilon^{2}} + Z^{\epsilon}(T,X)
	\end{equation}
for some scaled forcing term $Z^{\epsilon}(T,X)$. Note that eqn.~\eqref{eq:burgers2} is a discrete stochastic version of the Burgers' equation~\eqref{burgers}. Our computation lack rigor since we avoided any serious discussion regarding the random forcing terms and the corresponding scaling limits. The reader interested in more details should read \cite{BerGia}.

	\begin{theorem}[Theorem B.1 \cite{BerGia}]\label{thm:weak}
	The family $\{u^{\epsilon}\}_{\epsilon>0}$ defined by eqn.~\eqref{eq:discrete2} is weakly convergent as $\epsilon \rightarrow 0$ and the limit coincides with the mild solution of the stochastic Burgers' equation \eqref{burgers} if
	\begin{itemize}
	\item[(i)] $\sqrt{\epsilon} h^{ASEP}(0, \epsilon^{-1} x)$ converges weakly to a continuous function as $\epsilon \rightarrow 0$;
	\item[(ii)] for each $n \in \mathbb{N}$ there exist $a, c >0$ so that 
	\begin{equation}
	\sup_{x \in \mathbb{Z} + \frac{1}{2}} e^{- a \epsilon |x|} \mathbb{E}\left(e^{-n \sqrt{\epsilon} h^{ASEP}(0, x)}\right) \leq c;
	\end{equation}
	and
	\item[(iii)] for each $n \in \mathbb{N}$ there exist $a', c' >0$ so that 
	\begin{equation}
	\mathbb{E}\left(\epsilon^{n} (h(0, x) - h(0, y))^{2n}\right) \leq c' e^{a' \epsilon (|x| + |y|)}(\epsilon |x-y|)^n
	\end{equation}
for every $x, y \in \mathbb{Z}$ and all $\epsilon >0$.
	\end{itemize}
That is,
	\begin{equation}
	u^{\epsilon} \Rightarrow u.
	\end{equation}
	\end{theorem}

	\begin{rem}
	In Theorem \ref{thm:weak}, the initial conditions $h(0, x)$ may be non-deterministic, and hence, we consider the expected value for the growth conditions. Also, if $h(0,x)$ is flat or is a two-sided Brownian motion, then the growth conditions are satisfied, making Theorem \ref{thm:weak} applicable.
	\end{rem}

\subsection{Weak Asymmetric Limit Statistics}
\label{sec:WAL}

	In \cite{ACQ}, the authors use a scaling limit similar to the limit in Theorem~\ref{thm:weak} to obtain the probability distribution for the height function $h(t,x)$ of the KPZ equation \eqref{kpz} given that the \emph{wedge initial conditions}
	\begin{equation}
	h^{ASEP}(0, x) = |x|.
	\end{equation}
Note that, under the scaling in Theorem~\ref{thm:weak}, we have 
	\begin{equation}\label{eq:wedge_initial}
	\lim_{\epsilon \rightarrow 0} \sqrt{\epsilon} h^{ASEP}(0, x) = \begin{cases} \infty, \quad x \neq 0 \\ 0, \quad x=0 \end{cases}.
	\end{equation}
These initial condition are the so-called \emph{infinite narrow wedge initial conditions}, which may be interpreted as $h(0, x) = - \log(\delta_{x=0})$ after the scaling. Moreover, the authors in \cite{ACQ} consider the infinite wedge initial conditions since the solution to the KPZ equation, after the Cole-Hopf transformation with such initial conditions, gives the point-to-point partition function for a directed random polymer (see \cite{ACQ}). The work in \cite{ACQ} relies on the results of Bertin and Giacomin \cite{BerGia} and Tracy and Widom \cite{TW,TW09}. Note that the Theorem~\ref{thm:weak} may not be applied directly to the wedge initial conditions \ref{eq:wedge_initial}. Instead, the Amir, Corwin and Quastel consider the discrete height function at some (arbitrary) small finite time and show that then the height function satisfies the conditions in Theorem~\ref{thm:weak}. Thus, after an adequate limit, Amir, Corwin and Quastel are able to extend the result of Bertini and Giacomin \cite{BerGia} for the wedge initial conditions, and we obtain a scaling limit of the ASEP with wedge initial conditions to the the KPZ equation with infinite wedge initial conditions, as in Theorem \ref{thm:weak}. Moreover, from the work of Tracy and Widom \cite{TW, TW09}, Amir, Corwin and Quastel were able to obtain the probability distribution for $h(t,x)$ through careful asymptotic analysis of the formulas introduced by Tracy and Widom. In particular, the author are able to analyze the statistics of the height function for the KPZ equation for large and small time regimes. In the small time regime, the statistics of the height function for the KPZ equation are Gaussian statistics. In the large time regime, the statistics of the height function for the KPZ equation are given by the Tracy-Widom distribution.\par

	\begin{theorem}[Cor.~1.3 and 1.6\cite{ACQ}]
	Let $Z(t,x)$ to be a solution of the stochastic heat equation~\eqref{heat} with delta initial conditions $Z(0,x) = \delta_0(x)$, and define $F(t,x)$ by the formula 
	\begin{equation}
	Z(t, x) = \frac{e^{-x^2/2t}}{\sqrt{2 \pi t}} \exp F(t,x).
	\end{equation} 
Then,
	\begin{equation}
	\lim_{t \rightarrow \infty} \mathbb{P} \left( \frac{F(t, t^{2/3}x) + t/24}{t^{1/3}} \leq s\right) = F_{GUE}(2^{1/3} s),
	\end{equation}
and
	\begin{equation}
	2^{1/2} \pi^{-1/4} t^{-1/4} F(t, t^{1/2}x)
	\end{equation}
converges in distribution to a standard Gaussian as $t \rightarrow 0^{+}$.
	\end{theorem}

\section{Discrete Models}
\label{sec:DM}
The asymmetric simple exclusion process (ASEP), introduced by Spitzer \cite{spitzer} in 1970, is the epitome for interacting particle systems. It is the simplest example of random walkers on a lattice acting under an exclusion rule - no more than two particles per site - with arbitrary drift. It has become essential in the study of KPZ equaiton due to its scaling limit \cite{ACQ} to the Kardar-Parisi-Zhang (KPZ) equation \cite{kpz}.

\subsection{ASEP}
\label{sec:asep}

The Asymmetric Simple Exclusion Process (ASEP) on a periodic lattice is a continuous Markov process of $N$ particles moving on a discrete ring of length $L$, each of which has an exponential clock with parameter $1$, the particle chooses with probability $p$ (resp.~$q = 1-p$) to jump right (resp.~left) when the clock is activated, and the jumps is performed if the new position is unoccupied. The configuration space is 
	\begin{equation}
	S = \{ (\bar{x}_1, \dots, \bar{x}_N )  \in (\mathbb{Z}/ L \mathbb{Z})^N \mid \bar{x}_i \neq_L \bar{x}_j \text{ for } i \neq j \}
	\end{equation}
with the vector of $(\bar{x}_1, \dots, \bar{x}_N )$ representing the location of $N$ particles in the lattice ring $\mathbb{Z}/L\mathbb{Z}$. Equivalently, we may describe the state space as the basis of a Hilbert space isomorphic to $(\mathbb{C}^2)^{\otimes L}$. Indeed, choose a basis of $\mathbb{C}^2$: a \lq\lq{}spin up\rq\rq{} vector $|\uparrow\rangle$ and a ``spin down" vector $|\downarrow\rangle$. Then, translate a particle configuration $\vec{x} \in S$ to a spin up, spin down configuration on $(\mathbb{C}^2)^{\otimes L}$ by the bijection
	\begin{equation}
	\vec{x} \mapsto |\vec{x}\rangle:= \bigotimes_{i=1}^L |e_i \rangle
	\end{equation}
with$|e_i\rangle = |\uparrow\rangle$ if $i \in \vec{x}$ and $|e_i\rangle = |\downarrow\rangle$ otherwise. Then, we define the Hilbert space $\mathcal{H} \subset (\mathbb{C}^2)^L$ as
	\begin{equation}
	\mathcal{H}_{N,L} := \text{span}\{|\vec{x}\rangle \mid \vec{x} \in S \},
	\end{equation}
the vector space spanned by all $N$ particle configurations on the ring of length $L$. We write the infinitesimal generator of the continuous-time Markov process as a sum of local operators on $\mathcal{H}_{N,L}$. That is, the infinitesimal generator is $A = A_1 + \cdots  + A_N$ with the operator $A_i$ acting as the identity operator on all factors of $\mathbb{C}^2$ in the tensor product $(\mathbb{C}^2)^{L\otimes }$ except for the $i^{th}$ and $(i+1)^{th}$ factors (modulo $L$ on the index $i$) with the operator acting as
	\begin{equation}\label{eq:local}
	\left[\begin{array}{cccc}
	0 & 0 & 0 & 0 \\
	0 & -p & p & 0 \\
	0 & q & -q & 0 \\
	0 & 0 & 0 & 0 \\
	\end{array}\right]
	\end{equation}
on the basis $\{ |\uparrow \rangle \otimes |\uparrow \rangle, |\uparrow \rangle \otimes |\downarrow \rangle, |\downarrow \rangle \otimes |\uparrow \rangle ,|\downarrow\rangle\otimes |\downarrow\rangle\}$. Note that the operator $A$ is actually defined on $(\mathbb{C}^2)^{\otimes L}$, but in fact, the operator $A$ preserves the Hilbert space $\mathcal{H}_{N,L}$ (i.e.~$A \mathcal{H}_{N,L} \subset \mathcal{H}_{N,L}$). We considered $A$ as an operator on $\mathcal{H}_{N,L}$. We introduced the Hilbert space formalism for ease of notation in describing the dynamics, but we also use the particle configuration notation in the following arguments.\par

	Take the universal cover of the integer lattice (i.e.~$\pi: \mathbb{Z} \rightarrow \mathbb{Z}/L \mathbb{Z}$) and lift the particle configurations and dynamics of the periodic ASEP (i.e.~the infinitesimal generator $A$) to the integer lattice. Thus, we lift the configuration space $S$ and obtain the configuration space in the infinite integer lattice,
	\begin{equation}\label{eq:conf}
	\mathcal{X}_N(L) = \{ (x_1, \dots, x_N) \in \mathbb{Z}^N \mid x_1 < \cdots < x_N< x_1 + L \}. 
	\end{equation}
Note that the particles in $\mathcal{X}_N(L)$ are ordered, whereas in $S$, we did not make such a distinction. This ordering is introduced in order to keep track of the periodic condition succinctly and it is introduced by making an arbitrary choice of the rightmost particle on the initial conditions and pulling-back the particle configuration to the integer lattice according to a cyclic ordering. In any case, one may obtain any other ordering of the particles by a cyclic permutation of the particles and the overall dynamics will remain the same. Moreover, we may also lift the infinitesimal generator of the Markov process and the Hilbert space. The corresponding Hilbert space will be the $\ell^2$ vectors on an orthogonal basis $\{ | \vec{x} \rangle \mid \vec{x} \in \mathcal{X}_N(L) \}$,
	\begin{equation}
\mathcal{H} = \left\{ \sum_{  \vec{x} \in \mathcal{X}_N(L)  } c_{\vec{x}}  |  \vec{x} \rangle \hspace{1mm}  \bigg|  \hspace{1mm} \sum_{\vec{x} \in \mathcal{X}_N(L)} c_{\vec{x}}^2 < \infty \right\},
	\end{equation}
and the pull-back of the infinitesimal generator (which we denote as $A$ by an abuse of notation) may also be written as a sum of local operators that give the interaction of neighboring sites. Then, the evolution of the probability function of the Markov process is given by the \textit{Kolmogorov forward master equation}
\begin{equation}\label{kolmogorov}
\frac{d}{dt} \left( \sum_{\vec{x} \in \mathcal{X}_N(L)} \mathbb{P}_{\vec{y}}(\vec{x};t) |\vec{x} \rangle \right)= A^{t} \left( \sum_{\vec{x} \in \mathcal{X}_N(L)} \mathbb{P}_{\vec{y}}(\vec{x};t) |\vec{x}\rangle \right) 
\end{equation}
with $A^{t}$ is the transpose of $A$ and $\vec{y} \in \mathcal{X}_N(L)$ is an arbitrary initial particle configuration at time $t=0$. In the following section \ref{sec:Plancherel}, we describe the eigenvectors of the ASEP generator $A$, effectively diagonalizing the operator $A$. In the case that $L \rightarrow \infty$, we give exact formulas for joint probability function $\mathbb{P}_{\vec{y}}(\vec{x};t)$ given by a nested sequence of contour integrals in the complex plane (see Thm.~\ref{thm:TW}).

\subsection{Bethe Ansatz and Plancherel Formulas}
\label{sec:Plancherel}

	For the infinitesimal generator $A$ defined locally by eqn.~\eqref{eq:local} on the configuration space given by $\mathcal{X}_N(L)$ (see \eqref{eq:conf}), we find eigenvectors by the \emph{Bethe ansatz}. The Bethe ansatz was first employed by H. Bethe in \cite{Bethe} to diagonalize the operator for the one-dimensional Heisenberg model in 1931. Since then, the Bethe ansatz has been used in a myriad of (one-)dimensional interacting models (see \cite{Sutherland}). The ansatz takes solutions of the ``free" particles system (i.e.~the same particle systems without any interactions) and assumes that a solution of the interacting particle system may be written as a linear combination of solutions of the ``free" particle system. The coefficients of the linear combination of ``free" solutions are determined by the interaction of the particles. In general for the Bethe ansatz, it is not clear that that all of the solution of the interacting particle system are given through linear combinations of ``free" solutions. In the following, we apply the Bethe ansatz for ASEP on the ring of length $L \in \mathbb{N}$ and $N<L$ particles. In the case of the infinite lattice line (i.e.~$L \rightarrow \infty$), we can write the probability function of the ASEP as a linear combination of ``free" solutions through the \emph{Plancherel formulas} first established for ASEP by Tracy and Widom in \cite{TW} and generalized by Borodin, Corwin, Petrov and Sasamoto in \cite{BorodinCorwinPetrovSasamoto}.\par

	We decompose the equation~\eqref{kolmogorov} to obtain equations of ``free" (i.e.~non-interacting) particles. First, we rewrite~\eqref{kolmogorov} into an equation on the coefficients of the vectors. That is,
	\begin{equation}\label{master}
	\frac{d}{dt} \mathbb{P}_{\vec{y}} (\vec{x};t) = \sum_{i=1}^N [p  \mathbb{P}_{\vec{y}} (\vec{x}_i^-;t) \gamma_{i,i-1} + q  \mathbb{P}_{\vec{y}} (\vec{x}_i^+;t)\gamma_{i+1,i} - p  \mathbb{P}_{\vec{y}} (\vec{x};t)\gamma_{i+1,i} - q  \mathbb{P}_{\vec{y}} (\vec{x};t) \gamma_{i,i-1}],
	\end{equation}
with $\vec{x}_i^{\pm}$ denoting the particle configuration $(x_1, \dots, x_{i-1}, x_i \pm 1, x_{i+1}, \dots, x_N)$ and $\gamma_{i,j} := \mathds{1}(x_i \neq x_j +1)$ an indicator function denoting that the $i^{th}$ and $j^{th}$ particles are not adjacent. We further decompose this equation by using the \emph{coordinate Bethe ansatz} approach. We write eqn.~\eqref{master} as a \textit{free equation}~\eqref{free}, ignoring the interaction among particles, and set the boundary conditions~\eqref{boundary}, which encode the interaction of the particles. The free equations is simply given by taking all of the indicator functions $\gamma_{i, j}$ to be one,
	\begin{equation}\label{free}
	\frac{d}{dt} \mathbb{P}_{\vec{y}} (\vec{x};t) = \sum_{i =1}^N p \mathbb{P}_{\vec{y}} (\vec{x}_i^-;t) + q \mathbb{P}_{\vec{y}} (\vec{x}_i^+;t) - \mathbb{P}_{\vec{y}} (\vec{x};t),
	\end{equation}
and the boundary conditions are
	\begin{equation}\label{boundary}
	\mathbb{P}_{\vec{y}} (\vec{x};t) = p \mathbb{P}_{\vec{y}} (\vec{x}_i^-;t) + q \mathbb{P}_{\vec{y}} (\vec{x}_{i-1}^+;t)
	\end{equation}
for all $\vec{x} \in \mathcal{X}_N(L)$ such that $x_i = x_{i-1}+1$. We also have a \emph{periodic condition}
	\begin{equation}\label{periodic}
	\mathbb{P}(x_1, x_2, \dots, x_N;t) = \mathbb{P}(x_2, \dots, x_N ,x_1 +L;t ).
	\end{equation}
Then, we consider the \emph{eigenvalue problem}:
	\begin{equation}
	E  \mathbb{P}_{\vec{y}} (\vec{x};t) =  \sum_{i =1}^N p \mathbb{P}_{\vec{y}} (\vec{x}_i^-;t) + q \mathbb{P}_{\vec{y}} (\vec{x}_i^+;t) - \mathbb{P}_{\vec{y}} (\vec{x};t).
	\end{equation}
One may check that a solution to the eigenvalue problem (disregarding the boundary and periodic conditions) is given by
	\begin{equation}\label{freesoln}
	u(\vec{x}; t) = \sum_{\sigma \in S_N} A_{\sigma}(\vec{z}) \prod_{i=1}^N z_{i}^{x_{\sigma (i)}}
	\end{equation}
with $S_N$ the symmetric group on $N$ numbers and $\vec{z} = (z_1, \dots, z_N)$ a vector of $N$ (still undetermined) complex numbers. Then, we require that the \emph{free solution}~\eqref{freesoln} satisfies the boundary condition~\eqref{boundary} and the periodic condition~\eqref{periodic}. In terms of the coefficients of the free solution~\eqref{freesoln}, we must have that
	\begin{equation}
	A_{\sigma}(\vec{z}) = \prod_{\substack{i < j \\ \sigma (i) > \sigma(j) } } - \frac{p + q z_{\sigma (j)} z_{\sigma (i)} - z_{\sigma (i)}}{p + q z_{\sigma (j)} z_{\sigma (i)} - z_{\sigma (i)} },
	\end{equation}
	and
	\begin{equation}\label{Bethe}
z_j^{L} = (-1)^{N-1} \prod_{i=1}^N \frac{p +q z_i z_j - z_j}{p + q z_j z_i - z_i} \hspace{5mm} \text{for} \hspace{5mm} j= 1, \dots, N,
	\end{equation}
in order to satisfy the boundary and periodic conditions. Therefore, a solution to the so-called \emph{Bethe equations}~\eqref{Bethe} gives an eigen-solutions to the Komogorov's master equation~\eqref{kolmogorov}. It turns out that all eigen-solutions of~\eqref{kolmogorov} are of the form~\eqref{freesoln} with $\vec{z}$ satisfies the Bethe equations~\eqref{Bethe} (see \cite{Saenz}).\par

	We have solved for the eigen-solutions of the Kolmogorov master equation~\eqref{kolmogorov}, but we still want solutions that give probability functions. In particular, we want solutions to the eqn.~\eqref{kolmogorov} with the initial conditions
	\begin{equation}\label{eq:ICdelta}
	 \mathbb{P}_{\vec{y}} (\vec{x};0) = \delta_{\vec{y}}(\vec{x}) = \begin{cases} 1 \hspace{5mm} \vec{y} = \vec{x} \\ 0 \hspace{5mm}  \vec{y} \neq \vec{x} \end{cases}.
	\end{equation}
It turns out that the eigen-solutions~\eqref{freesoln} given by the Bethe ansatz form a basis of the Hilbert space of configurations $\mathcal{H}_{N, L}$ (see \cite{Saenz}), which we call the \emph{Bethe basis}. Also, note that the delta functions $\delta_{\vec{y}}$, for all configurations $\vec{y} \in \mathcal{X}_N(L)$, make up a basis for $\mathcal{H}_{N, L}$, which we call the \emph{configuration basis}. Therefore, it suffices to give a linear transformation from the Bethe basis to the configuration basis to solve the Kolmogorov master equation~\eqref{kolmogorov} with delta initial conditions~\eqref{eq:ICdelta}. A linear transformation (with certain restrictions on the inner-product) between the basis is called a \emph{Plancherel formula} (see \cite{BorodinCorwinPetrovSasamoto}). \par

	\begin{theorem}[C.~Tracy and H.~Widom \cite{TW}]\label{thm:TW}
	For the ASEP with $N$ particles on $\mathbb{Z}$ so that the jumping rate to the left is $p$ and the jumping rate to the right is $q:= 1-p$, the probability function with initial conditions $\vec{y} =(y_1 < y_2< \cdots <y_N)$  at time $t = 0$ is given by the sum of nested contour integrals
	\begin{equation}\label{eq:prob_fun}
	\mathbb{P}_{\vec{y}}(\vec{x}; t)  = \frac{1}{(2 \pi i)^N} \sum_{ \sigma \in S_N}\oint_C d \xi_1 \cdots \oint_C d \xi_N A_{\sigma}(\vec{\xi}) \prod_{j=1}^N \left[ \xi_{\sigma(j)}^{x_j - y _{\sigma(j)} -1} e^{\epsilon(\xi_j)t}\right].
	\end{equation}
for any $\vec{x} = (x_1< x_2 < \cdots < x_N)$, $\epsilon (\xi) := p \xi^{-1} + q \xi -1$, and small enough simple closed contours $C$ containing the origin inside.
	\end{theorem}

	In \cite{TW}, Tracy and Widom give a formula for $\mathbb{P}_{\vec{y}} (\vec{x};0)$ by using the Bethe ansatz and using a Plancherel formulas for the case $L \rightarrow \infty$ and $N = L/2$. At the technical level, taking the limit $L \rightarrow \infty$ with $N = L/2$ makes the Bethe equations~\eqref{Bethe} trivial. In general, finding the solutions of the Bethe equations~\eqref{Bethe} explicitly is non-trivial. Thus, results similar to those by Tracy and Widom in \cite{TW} are less accessible for finite $L$, but there has been some significant progress for the case with $p =1$ (see \cite{baikliu2016, baikliu2018}). In \cite{BorodinCorwinPetrovSasamoto}, Borodin, Corwin, Petrov, and Sasamoto give Plancharel formulas for other particle systems related to ASEP that also behave well under the Bethe ansatz.

\subsection{Scaling Exponents}
\label{sec:scaling}

	The existence of the invariant measure for the KPZ equation is non-trivial~\cite{FunkaiQuastel}, and, equivalently, one might find the corresponding result for the stochastic heat equation in \cite{Tess}. It turns out that the invariant measure is Brownian motion up to a constant time dependent height shift \cite{SasamotoSpohn}. This may be motivated by the ASEP, which serves as a discrete approximation to the KPZ equation according to section~\ref{sec:Bertini}. For the periodic ASEP, we determine the invariant measure below. Heuristically, this gives you an invariant measure for the KPZ equation~\eqref{kpz} by taking a scaling limit as in Theorem \ref{thm:weak}, and this allows for further heuristic arguments that determine the KPZ 1:2:3-scaling.\par

	For ASEP with only one particle, it is clear that the process is equivalent to a random walker with some drift determined by the asymmetry of the ASEP. In this case, the invariant measure is given by having each location for the single particle to be equally likely. In general, for ASEP with a fixed deterministic number of particles, the invariant measure will be uniform on the possible configurations. This may be easily checked by letting $\mathbb{P}(\vec{x} ; t) =const$ in the Kolmogorov forward master equation~\eqref{kolmogorov}, which describes the evolution of the ASEP. Then, for ASEP without conditions on the number of particles, we construct an invariant measure so that each location $x \in  \mathbb{Z} + 1/2$ is occupied based on an independent Bernoulli variable $B_x$ with parameter $b \in [0,1]$, and we denote this measure by $P_s$. More specifically, for the occupation variables $\eta_t(x) : = \mathds{1}(x \in \vec{x}(t))$, we have
	\begin{equation}\label{eq:stationary}
	P_s(\eta_t(x_1)=1, \dots, \eta_t(x_k)=1 ) = b^{k}
	\end{equation}
for $k \in \mathbb{N}$, $x_i \in \mathbb{Z} +1/2$, and $x_i \neq x_j$ if $i \neq j$. Then, it is not difficult to check that 
	\begin{equation}
	P_s (\vec{x} ; t) = P(\vec{x}'; t) \hspace{5mm} \text{if $\vec{x}$ and $\vec{x}'$ have the same number of particles}.
	\end{equation}
Moreover, we have that the Kolmogorov forward master equation~\eqref{kolmogorov} preservers the number of particles in a configuration as the process evolves. Therefore, we have that the probability function~\eqref{eq:stationary} gives a stationary measure for ASEP for any $b \in [0,1]$.\par

	The height function $h^{ASEP}(t, k)$ is given as a sum of the occupation variables $\eta_t(s)$, see eqn.~\eqref{eq:ASEP_height}. Then, for the stationary measure $P_s$, it is clear that the height function $h^{ASEP}(t,x)$ will be a two-sided random walk at any fixed time $t$. For $b \in (0,1)$, the height function $h^{ASEP}(t,x)$ will experience a deterministic linear growth depending on $b$. So, the stationary distribution for the height function $h^{ASEP}(t,x)$ will be a double-sided random walk plus a deterministic linear shift. In particular, via a scaling similar to the that of Theorem \ref{thm:weak}, we have that the invariant measure of the KPZ height function $h(t,x)$ will be a two-side Brownian motion plus a deterministic linear sift \cite{BerGia}.\par
 
	Now, we assume that the spatial variable $x$ in the stationary height function $h(t, x)$ scales according to the Gaussian scaling:
	\begin{equation}\label{eq:statScaling}
	h(x,t) \overset{D}{=} \varepsilon^{1/2} h( \varepsilon^{-1} x, t)
	\end{equation}
with equality in distribution and $\epsilon >0$. We ignore the deterministic height shift for simplicity. Then, we use this information to determine the KPZ 1:2:3-scaling.\par

	Take the height function $h(x,t)$ from the KPZ equation~\eqref{kpz} and consider the scaling
\begin{equation}
h_{\varepsilon}(t,x) := \varepsilon^{c} h (\varepsilon^{-b}t, \varepsilon^{-a}x).
\end{equation}
Then, the scaled height function $h_{\varepsilon}(t,x)$ satisfies a scaled version of the KPZ equation:
	\begin{equation}
	\partial_t h_{\varepsilon} = \frac{\varepsilon^{2a-b}}{2} \partial^2_x h_{\varepsilon} - \frac{\varepsilon^{2a-c-b}}{2} \left( \partial_x h_{\varepsilon} \right)^2 +\varepsilon^{\frac{a-b}{2}+c} W.
	\end{equation}
Furthermore, assume that we take the stationary height function $h(x, t)$ with the scaling~\eqref{eq:statScaling}. Thus, we set $a=1$ and $c=1/2$. Then, one considers the limit $\varepsilon \rightarrow 0$ and notes that, in order to keep the non-linear term in the KPZ equation for the scaled height function from blowing up to infinity or from vanishing to zero, one should set $b=3/2$. That is, we set
\begin{equation}\label{123}
h_{\varepsilon}(t,x) = \varepsilon^{1/2} h (\varepsilon^{-3/2} t, \varepsilon^{-1} x ),
\end{equation}
and we obtain the KPZ 1:2:3-scaling.

\subsection{Asymptotic Analysis}
\label{sec:Asymptotic}

	There are many models (e.g.~ASEP, directed polymers, stochastic higher spin six vertex model, polynuclear growth, etc) that belong to the KPZ universality class. All of these examples have a notion of a height function (or a Cole-Hopf transformation of a height function). More importantly, the height function for all these models have been shown to have the KPZ scaling~\eqref{123} (see \cite{Corwin, CorPet, Quastel}). This scaling behavior can be taken to be a practical weak definition of the \textit{KPZ universality class}. Additionally, when accessible for models in the KPZ universality class, the limiting distribution for the height function under the KPZ scaling~\eqref{123} is given by the \emph{Airy line ensemble} or one of its relatives. As such, one should expect a stronger criteria for models in the KPZ universality class that considers the limiting distribution of a height function under the KPZ scaling~\eqref{123}. In the case of determinantal process, limiting distribution results are more accessible (see \cite{okounkov03,borodin16}). 
\begin{theorem}[Theorem 1.1 in \cite{PraSpo}]\label{Airy}
Take $h(t, x)$ to be the height function for the polynuclear growth model and let $A(x)$ be the stationary Airy process. Then in the sense of weak convergence of finite dimensional distributions
	\begin{equation}
\lim_{t \rightarrow \infty} \frac{h(t, x t^2/3) - 2 t}{t^{1/3}} = A(x) - x^2
	\end{equation}
\end{theorem}
The \emph{Airy process} was introduced by M. Praehofer and H. Spohn in 2002 \cite{PraSpo}. Roughly speaking, the Airy process describes the edge path of Dyson's Brownian motion, which describes the dynamics of the eigenvalues of a Hermitian matrix with the entries of the matrix evolving as independent Brownian motions (see \cite{dyson62}). Similar limiting results have also been proven for the discrete polynuclear growth model, the north polar region of the Aztec diamond, and TASEP (see \cite{Shinault} for references). In particular, we note that the one point distribution function of the Airy process is given by the Tracy-Widom GUE distribution, $F_2(s)$, which describes the distribution of the largest eigenvalue of an $n \times n$ matrix under the Gaussian unitary ensemble as $n$ goes to infinity (see sec.~\ref{sec:TW}). That is,
	\begin{equation}
	\mathbb{P}[A(x) \leq s] = F_2(s).
	\end{equation}
In fact, the Tracy-Widom distribution (and its relatives) is more attainable for a wider class of models meaning that proving the explicit limit of the probability distribution of the height function at a point converges to the Tracy-Widom distribution is more attainable. For instance, in the case of the ASEP, which does not have a clear determinantal structure such as the TASEP, we have 
\begin{theorem}[Theorem in \cite{TW09}]\label{F2}
For ASEP with initial conditions $\vec{y}= (-1, -2, \dots)$, the one point function for the height function under the KPZ scaling is given by the Tracy-Widom distribution:
	\begin{equation}
\lim_{t \rightarrow \infty} \mathbb{P}\left( \frac{h(\frac{t}{q-p} , 0) - t/4}{2^{-4/3} t^{1/3}}  \geq -s \right) = F_2(s).
	\end{equation}
\end{theorem}
The ASEP was discussed in detail in section \ref{sec:asep}. These type of result of convergence to the Tracy-Widom distribution have been proven for other models that are believed to be in the KPZ universality class such as polynuclear growth model, last passage percolation, stochastic higher spin six vertex model, and directed polymers in a random environment (see \cite{Corwin, CorPet, Quastel, oconnell12, oconnell14, saenz2018}).\par

\begin{figure}[h]
\centering
\includegraphics[width=0.5\textwidth]{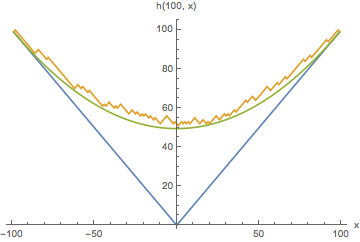}
\caption{A simulation of the ASEP with initial conditions $\vec{y} = (-1, -2, \dots)$ compared to the limiting shape $ \frac{t}{2} + \frac{x^2}{2t} $ for $|x|<t$.}
\end{figure}

	There are many layers to the KPZ universality class. In its most refined form, one would wish to show that a model belongs to the KPZ universality class by showing that the limiting process is the KPZ fixed point from \cite{matetski2016}, but this is a non-trivial task for most models that are believed to be in the KPZ universality class. In practice, we have two basic checks for some stochastic process to belong to the KPZ universality class: the scaling of the height function and the limiting distribution of the height function. One uses the KPZ 1:2:3-scaling~\eqref{eq:scalingKPZ} as a litmus test, and then one attempts to extract the GUE Tracy-Widom distribution (or its relatives, depending on the initial conditions). The KPZ universality class has gained much popularity in the last decade because many models that pass these checks have been discovered leading to deeper insight into the KPZ universality class.\par

	We have left out many important details out of the discussion of the KPZ universality class in exchange for brevity. For the interested reader, there are many survey that include far more details with\cite{Corwin, Quastel} particularly enlightening.

\section{Random Matrix Theory}
\label{sec:RMT}

	Random matrix theory has been introduced many times in different setting: by Hurwitz in 1890's for representation theory \cite{hurwitz1897,hurwitz1898}, by Wishart in 1928 for statistics \cite{wishart28}, by Wigner in the 1950s for quantum mechanics \cite{wigner57, porter65}, and by Robert May in the 1970's for evolutionary biology \cite{may72}. See \cite{diaconis17} for more on the history of random matrices.  In quantum mechanics, the idea is to consider a large quantum system with most of the ``quantum numbers" (except for spin and parity) averaged out and apply ideas of statistical mechanics to obtain macroscopic observables. For instance, we will consider a system of electrons on a line and we will obtain the distribution of the particles on the line as the number of particles increases to infinity. In this example, we encounter one of the most well-known and celebrated results in random mamtrix theory: Theorem~\ref{wigner}, the Wigner semi-circle law.\par 

	First, let us establish our model: electron on a continuous infinite line. Mathematically, this is a configuration of $n$ particles on the real line $\{(x_1, \dots, x_n) | x_i \in \mathbb{R} \}$ where each configuration has certain \textit{energy} (or weight) given by the \textit{Hamiltonian} function:
	\begin{equation}\label{eq:ham}
	H(x_1, \dots, x_n) := - \sum_{1 \leq i < j \leq n} \log |x_i - x_j|.
	\end{equation}
The \emph{thermodynamic equilibrium} is the configuration that minimizes the Hamiltonian function~\eqref{eq:ham} (i.e.~the energy of the configuration), and we are interested in determining the thermodynamc equilibrium for the electrons on a line. The avid reader will note that Hamiltonian function~\eqref{eq:ham} will force the electrons to repel each other with the lowest energy configuration given by all the electrons going off to infinity. Thus, in order to have a bounded thermodynamical equilibrium, we introduce an external  square potential that forces the electrons together giving us the Hamiltonian,
	\begin{equation}\label{eq:ham2}
	H(x_1, \dots, x_n) := \alpha \sum_{i=1}^n x_i^2- \sum_{1 \leq i < j \leq n} \log |x_i - x_j|, \hspace{5mm} \alpha >0.
	\end{equation}
This model with the Hamiltonian function \eqref{eq:ham2} is called the \textit{Coulomb gas model with $\log$-potential} (see \cite{Dyson}) and we will stick to this model for the rest of the section and apply methods of random matrix theory to find the distribution of the particles on the line. \par

	We set up the measure of our configurations of this model. Recall that in statistical mechanics (see \cite{Baxter} for an elementary introduction), one aims to study the \emph{partition function} of a model (i.e.~a generating function for a model) defined by
\begin{equation}\label{partition}
Z := \int_{\mathbb{R}^n} e^{-\frac{H(\vec{x})}{kT}} \mu(dx_1) \cdots \mu(dx_n)
\end{equation}
where the integral is over the space of all configurations with an appropriate measure $\mu$ (many times this will be clear form the model and other times this will be a delicate issue), $H$ is the Hamiltonian function of the model, $k$ is the Boltzmann constant, and $T$ is temperature. In the mathematics literature, the Boltzmann constant is set to be one and $T^{-1}$ is replaced by the parameter $\beta$. In particular, this defines a measure in the configuration space where the probability (density) of a configuration $\vec{x} = (x_1, \dots, x_n) \in \mathbb{R}^n$ is given by
\begin{equation}\label{probC}
\mathbb{P}(\vec{x}) = \frac{1}{Z} e^{- \beta H(\vec{x}) }
\end{equation}
In the following, we consider the distribution of the particles as $n$ goes to infinity. To be more precise, we consider the \textit{empirical distribution measure}
\[
\mu_{ED,n} = \frac{1}{n} \sum_{i=1}^n \delta_{x_i},
\]
scaled to have volume one, as $n \rightarrow \infty$.

\begin{rem}
We note that in the definition of the Coulomb gas model, we have two arbitrary constant: $\alpha$ the strength of the square potential, and $\beta$ the inverse temperature. Of course, the distribution of the particles vary as the parameters do. A strong square potential (i.e. $\alpha \gg 0$) will make the particles concentrate near the origin, and a low temperature (i.e. $\beta \gg 0$) will make the particles spread out. This becomes particularly delicate when the number of particles increases. In the following, we will be able to see the proper parameters that will give a non-trivial distribution as $n$ goes to infinity.
\end{rem}

\subsection{Wigner Semi-circle Law}
\label{sec:wigner}

	Before we take a step forward in solving the Coulomb gas model, we take a step back to consider random matrices. The introduction of random matrices in this section may seem deus ex machina, and this seems to be the theme of integrability. The point is that the following results are attainable due to the strong structure that is hidden in the background of the equations. By introducing random matrices, we will see that the partition function~\eqref{partition} will appear naturally and the computations needed will be (more) straight forward.\par

	Define our ensemble of random matrices as Hermitian matrices each with entries that have a normal distribution $\mathcal{N}(\mu , \sigma^2)_{\mathbb{C}}$. For convenience, we normalize our matrices so that $\mu = 0 $ and $\sigma^2 = 1$ by scaling and centering arguments. Also, we would like our entries to be independent, but that is not fully possible if we want to keep the Hermitian structure (i.e. $M_{i,j} = \bar{M}_{j,i}$). So, we take a random Hermitian matrix defined by $M = (M_{i,j})_{i,j=1}^n$  with $\text{Re }M_{i,j} \sim \mathcal{N}(0,1/2)_{\mathbb{R}}$ for $i<j$, $\text{Im }M_{i,j} \sim \mathcal{N}(0,1/2)_{\mathbb{R}}$ for $i<j$, $M_{i,j} = \bar{M}_{j,i}$ for $j<i$, and $M_{i,i} \sim \mathcal{N}(0,1)_{\mathbb{R}}$. This is the \textit{Gaussian Unitary Ensemble (GUE)}, and from this ensemble, we will compute the probability distribution of the eigenvalues of random matrix. The probability distribution of the GUE is given by the product of the probability distributions of the independent entries. That is,
	\begin{equation}
\mathbb{P}(dM) = \prod_{1 \leq i < j \leq n} \mathbb{P}(d \text{ Re} M_{i,j}) \prod_{1 \leq i < j \leq n} \mathbb{P}(d \text{ Im} M_{i,j}) \prod_{k=1}^n \mathbb{P}(dM_{k,k}).
	\end{equation}
Since each independent variable has a normal distribution, we may write the probability function with respect to the Lebesgue measure:
	\begin{equation}
	\mathbb{P}(dM) = \frac{1}{Z} \exp{\left( - \frac{1}{2} \text{Tr} M^2 \right)} \prod_{1 \leq i < j \leq n} d \text{ Re} M_{i,j} \prod_{1 \leq i < j \leq n} d \text{ Im} M_{i,j} \prod_{k=1}^n dM_{k,k},
	\end{equation}
for some normalizing constant $Z$. Note that the probability density function with respect to the Lebesgue measure (i.e.~$ \exp{- \frac{1}{2}  \text{Tr} M^2}/Z$) is invariant under unitary transformations. Moreover, we may diagonalize any Hermitian matrix by some unitary matrix. Say 
\begin{equation}\label{usub}
M = U D U^{-1}
\end{equation}
with $M$ a Hermitian matrix, $U$ some unitary matrix, and $D = (y_1, \dots, y_n)$ a diagonal matrix. Therefore, we use unitary matrices to define a change of variables $M \mapsto (U, D)$, and under this change of variables we obtain the marginal distribution of the eigenvalues $D$.\par

	Differentiating equation \eqref{usub}, we have that
	\begin{equation}
	d M  = dU D U^{-1} + U dD U^{-1} + U D d U^{-1}.
	\end{equation}
Recall that $d U^{-1}  = - U^{-1} dU U^{-1}$ since $d I = d [U U^{-1}] = dU U ^{-1} + U dU^{-1}$. Then,
	\begin{equation}
	\begin{split}
	U^{-1} dM U  &= U^{-1} dU D - D U^{-1} dU + d D\\
	& = [U dU^{-1}, D ]  + dD.
	\end{split}
	\end{equation}
In coordinate form, we have
	\begin{equation}\label{1forms}
	\left[ U^{-1} dM U \right]_{i,j}= \sum_{k=1}^n (U^{-1})_{k,i} dU_{k,j}(y_j - y_i) + \delta_{i,j} dy_{i}.
	\end{equation}
This formula is enough to compute the Jacobian $J_{M}^{(U,D)}$ for the change of variables $M \mapsto (U, D)$. In particular, we have
	\begin{equation}
	\det J_{M}^{(U,D)} = f(U) \prod_{j <k} (y_j -y_k)^2 
	\end{equation}
for some function $f$ that is independent of the $y$-variables.
Therefore, we can use~\eqref{usub} as a u-substitution to write the probability density of the GUE and also use~\eqref{1forms} to write the Jacobian of the u-substitution. Then, after integrating out the terms depending on $U$ (or taking the marginal distribution of $M = (U, D)$ with fixed eigenvalue configuration $D$), one has the classical result:
\begin{theorem}[Theorem 3.3.1 \cite{Mehta}]
The joint probability density function for the eigenvalues of matrices for the Gaussian Unitary Ensemble is given by
\begin{equation}\label{probM}
P_{n} (y_1, \dots, y_n) = C_{n} \exp{\left(- \frac{1}{2} \sum_{i=1}^n y_i^2 + 2 \sum_{j<k} \log |y_j - y_k| \right)}.
\end{equation}
The constant $C_{n}$ is chosen in such a way that $P_{n}$ is normalized to unity:
\[
\int_{-\infty}^{\infty} \cdots \int_{-\infty}^{\infty} P_{n} (y_1, \dots, y_n) dy_1 \cdots dy_n =1.
\]
\end{theorem}
We have that the probability distribution of a given configuration of particles in the Coulomb gas model~\eqref{probC} is the same as the probability distribution of the eigenvalues of the GUE after identifying the variables $x_i = y_i$ for $i =1, \dots, n$ and setting the parameters to $\beta =2$ and $\alpha = 1/4$. This is now a leap forward in the understanding of the distribution of the particles in the Coulomb gas model. We will give a sketch of the proof of the following theorem in the section \ref{sec:stj}.
\begin{theorem}[Wirgner's semicircle law Theorem 2.4.2 \cite{Tao}]\label{law}
Let $M_n$ be an $n\times n$ GUE matrix and define the empirical spectral distribution (ESD) by $\mu_{n} = \frac{1}{n}\sum_{i=1}^n \delta_{y_i/\sqrt{n}}$, where $\{y_1, \dots, y_n\}$ are the eigenvalues of $M_n$. Then, the ESDs $\mu_{n} $ converges almost surely in distribution to the Wigner semi-circle distribution
\begin{equation}\label{wigner}
\mu_{sc} = \frac{1}{2 \pi} (4 - |x|^2 )^{1/2} \mathds{1}_{[-2,2]} dx.
\end{equation}
\end{theorem}
\begin{figure}[h]
\centering
\includegraphics[width=0.5\textwidth]{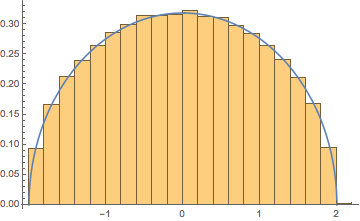}
\caption{Simulation of the distribution of a $500 \times 500$ GUE matrix sampled once. The smooth curve is the graph of the Wigner semicircle distribution.}
\end{figure}
\begin{rem}
Note that the position of the eigenvalues is scaled by $\sqrt{n}$. This result is somewhat unclear from the point of view of the Coulomb gas model, but it becomes more transparent from the point of view of random matrix theory. Indeed, we have that
\begin{align*}
\mathbb{E} \text{Tr} M_n &= \sum_{i=1}^n \mathbb{E} M_{i,i} =0\\
\mathbb{E} \text{Tr} M_n^2 &= \sum_{i=1}^n \mathbb{E} M_{i,i}^2 =n,
\end{align*}
which is the first sign that eigenvalues of $M_n$ are of order $n^{1/2}$ (since the trace of the matrix is the sum of the eigenvalues). This statement can be more precise showing the concentration of the eigenvalues of $M_n$ is between $-2 \sqrt{n}$ and $2 \sqrt{n}$. We invite the curious reader to check \cite{Tao} for reference.
\end{rem}

\subsection{Tracy-Widom Distribution}
\label{sec:TW}

	The \emph{Tracy-Widom distribution} was discovered in the work of Tracy and Widom in \cite{TW94} under the settings of random matrices. In the 60's and 70's, methods from inverse scattering theory such as Riemann-Hilbert problems or isomonodromic deformations were implemented to compute the local statistics of the random matrix ensembles such as the GUE. The Tracy-Widom distribution corresponds to the local statistics near the top eigenvalue in the GUE. In particular due to the work of Tracy and Widom \cite{TW94}, the the cumulative distribution function of the Tracy-Widom distribution, denoted by $F_{2}(s)$, is given by
	\begin{equation}\label{eq:TWdef}
	F_{2}(s) := \lim_{n \rightarrow \infty} P\left(\frac{y_{max} - \sqrt{2n}}{(2^{-3} n)^{1/6} } \leq s \right).
	\end{equation}\par

\begin{figure}[h]
\centering
\includegraphics[width=0.5\textwidth]{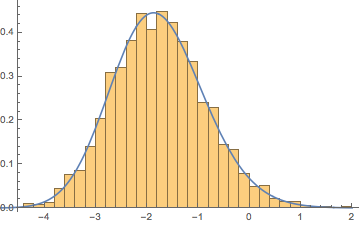}
\caption{Simulation of the distribution of the largest eigenvalue of a $100 \times 100$ matrix with a sample size of $4000$. The smooth curve is just smooth version of the bar graph. Mean of the simulation is $-1.74764$ with standard deviation $0.899264$, where the predicted values are $-1.771086807411$ and $0.8131947928329$.}
\label{TWdist}
\end{figure}

	The function $F_2(s)$ in eqn.~\eqref{eq:TWdef} may be written with explicit functions using a solution of the II-Painlev\'e equation. In particular, let $q(s)$ be a solution of the II-Painlev\'e equation,
	\begin{equation}
	\frac{d^2 q(s)}{ds} = s q(s) + 2 q(s)^3
	\end{equation}

\subsection{Stieltjes Transformation}
\label{sec:stj}

There are many ways to prove Wigner's semi-circle law, Theorem~\ref{wigner}. Here, we consider the \textit{Stieltjes transform}, which maps a measure on the real line to a complex function on a branch cut of the complex plane. For a measure $\mu$ on the real line $\mathbb{R}$, denote the Stieltjes transform with $s_{\mu}: \mathbb{C} \setminus \mathbb{R} \rightarrow \mathbb{C}$ defined by 
	\begin{equation}
	s_{\mu}(z) := \int_{\mathbb{R}} \frac{1}{x-z} \mu(dx).
	\end{equation}
If the measure $\mu$ relatively continuous to the Lebesgue measure (i.e.~ $\mu(dx) = p(x)dx$ for a continuous function $p(x)$), we can take the inverse of the Stietjes transform (see \cite{Tao} for reference). For instance, if $\mu(x) = p(x) dx$, then 
	\begin{equation}
	p(x) = \lim_{b \rightarrow 0^{+}}\frac{s_{\mu}(x + i b) - s_{\mu}(x - ib)}{2 \pi i}.
	\end{equation}
The proof of the Wigner's semicircle law follows by writing a recursive equation for the Stieltjes transform of the empirical distribution as follows
\begin{align*}\allowdisplaybreaks
\mathbb{E} s_{\mu_{n}}(z) &= \mathbb{E}\int_{\mathbb{R}} \frac{1}{x-z} \left( \frac{1}{n} \sum_{i=1}^n \delta_{\lambda_i / \sqrt{n}}\right)= \mathbb{E} \frac{1}{n} \sum_{i=1}^n \frac{1}{\lambda_i/ \sqrt{n} -z} \\
&= \mathbb{E} \text{Tr}\left(\frac{M}{\sqrt{n}} - z I\right)^{-1}= \mathbb{E} \sum_{i=1}^n \left[ \left( \frac{M}{\sqrt{n}} - z I \right)^{-1} \right]_{i,i}\\
&= \mathbb{E}  \left[ \left( \frac{M}{\sqrt{n}} - z I \right)^{-1} \right]_{n,n}
\end{align*}
At this point, we can use the \textit{Schur complement formula}:
\[
\left[
\left(
\begin{array}{cc}
A & B \\
C & d\\
\end{array}
\right)^{-1}
\right]_{n,n} = \frac{1}{d - C A^{-1} B}
\]

Therefore,
	\begin{equation}\label{eq:prelim}
	\begin{split}
	\mathbb{E} s_{\mu_n}(z) &= - \mathbb{E} \frac{1}{z + \frac{1}{n} C (\frac{1}{\sqrt{n}} M_{n-1} - z I) B}\\
		&= - \mathbb{E} \frac{1}{z + s_{n-1}(z)} + o(1).
	\end{split}
	\end{equation}
Now,  assume that $\mu_n \rightarrow \mu$ as $n \rightarrow \infty$ if and only if $s_{\mu_n} \rightarrow s_{\mu}$ as $n \rightarrow$ (see \cite{Tao} for technical details). Then, we have that $\mu_n \rightarrow \mu$ for $n \rightarrow \infty$ with 
	\begin{equation}\label{0cat}
	s_{\mu}(z) = \frac{-1}{z + s_{\mu}(z)}
	\end{equation}
by taking the limit of eqn.~\eqref{eq:prelim}. The functional equation~\eqref{0cat} has two solutions. One has to be careful to choose the right one, and then, one can take the inverse Stiltjes to obtain the semicircle law as desired.\par
\begin{rem}
There is a concentration result that allows us to do all the previous computations with the expected value and then revert back to the random variable.
\[
\mathbb{P} \left( |s_n(a +i b) - \mathbb{E} s_{n}(a +i b)| \geq \frac{\lambda}{\sqrt{N}}\right) \leq C e^{- \lambda^2}
\]
\end{rem}
The final result is very important, but the reason we sketched the proof is that in the process we see the equation~\eqref{0cat} emerge. This is the \textit{spectral curve} which plays a central role in the computations of the EOC topological recursion for the generalized Catalan numbers described in sec.~\ref{sec:EOC}.

\subsection{EOC Topological Recursion}
\label{sec:EOC}

	Let's keep working on the Coulomb gas model with $\beta =2$ and $\alpha = 1/4$ (i.e. the GUE ensemble), and consider further statistics on the position of the particles. In particular, we will now consider the distribution of the finding a $k$ points at $k$ given positions. We note that we have partilly answer this question for the case when $k=n$ in section \ref{wigner} by the joint probability density function for the eigenvalues of the GUE ensemble, which is given by equation~\eqref{probM}. Then, from taking the distribution of the $n$ particles at $n$ given points, it is not difficult to find the distribution of $k$ particles, which is done by integration of $n-k$ particles (or rather allowing $n-k$ particles to be anywhere). Thus, we introduce the \textit{$k$-point correlation function}:
\begin{equation}\label{kpoint}
W_k(y_1, \dots, y_k) := \frac{n!}{(n-k)!}\int_{- \infty}^{\infty} \dots \int_{-\infty}^{\infty} P_n(y_1, \dots, y_n) dy_{k+1} \cdots dy_{n},
\end{equation}
where $P_n$ is given by~\eqref{probM}.\par

	In this context, the Wigner semi-circle law, Theorem~\ref{law}, states that the $1$-point correlation function of the Coulomb gas model (for our chosen parameters) is given by the semicircle distribution as $n$ goes to infinity. Along this line of thought, it becomes natural to ask for an asymptotic expansion of the $k$-point correlation function. This asymptotic expansion is given by the \textit{EOC topological recursion} (defined below) named after B. Eynard, N. Orantin, and L. Chekhov for their work in \cite{chekhov06, eynard07}. Before we define the EOC topological recursion, we will work on our example of the Coulomb gas and perform some computations to motivate the general definition of the EOC topological recursion. \par
	First, we rewrite the right-hand-side of~\eqref{kpoint} using integrals in the space of $n \times n$ Hermitian matrices (i.e.~in the GUE ensemble) instead of the integrals in the space of the particle configurations. We perform this calculation explicitly for the one point function. Note the equalities
	\begin{align*}
	\int_{- \infty}^{\infty} \cdots \int_{-\infty}^{\infty} P_n(z, y_2, \dots, y_n)dy_2 \cdots dy_n &= \int_{- \infty}^{\infty} \cdots \int_{-\infty}^{\infty} P_n(y_1, \dots, y_n) \delta(z - y_1) dy_1 \cdots dy_n\\
	& \int_{- \infty}^{\infty} \cdots \int_{-\infty}^{\infty} P_n(y_1, \dots, y_n) \frac{1}{\pi (z-y_1)} dy_1 \cdots dy_n.
	\end{align*}
The first equality is straightforward from the definition of the delta function $\delta (x)$, but the second equality is given by a residue calculation on the complex plane. In particular for the second equality, one considers the integral on $dy_1$ to be a contour integral on the complex plane, deforms the contour around $y_1 =z$ by a ball of radius $\epsilon$, and then take the limit as $\epsilon$ goes to zero to obtain the result. Therefore, we can write the $1$-point function
	\begin{equation}
	W_k(z) = \int_{- \infty}^{\infty} \cdots \int_{-\infty}^{\infty} P_n(y_1, \dots, y_n) \sum_{i=1}^{n}\left( \frac{1}{\pi (z-y_i)} \right) dy_1 \cdots dy_n.
	\end{equation}
At this point, note that we can write the sum in the integral as follows
	\begin{equation}
	\text{Tr} \frac{1}{z  - M} = \sum_{i=1}^{n}\left( \frac{1}{\pi (z-y_i)} \right), 
	\end{equation}
with $M$ a $n \times n$ Hermitian matrix with eigenvalues $(y_1, \dots, y_n)$. For example, the 1-point function is given by
	\begin{equation}\label{eq:OnePoint}
	W_1(z) = \mathbb{E}\left[ \text{Tr} \frac{1}{z-M} \right]
	\end{equation}
with the expected value taken over the measure on the space of $n\times n$ Hermitian matrices given by $\exp{\left(-\text{Tr} M^2\right)} dM$. We use eqn.~\eqref{eq:OnePoint} to give an asymptotic expansion of the $1$-point function. We use a geometric series expansion of $(z-M)^{-1}$, and then
	\begin{equation}
	\begin{split}
	\mathbb{E}\left[ \text{Tr} \frac{1}{z-M} \right] &= \mathbb{E}\left[ \text{Tr} \sum_{j=0}^{\infty}z^{j-1} M^j \right] \\
	&= \sum_{j=0}^{\infty} z^{j-1} \mathbb{E}\left[ \text{Tr} M^j \right].
	\end{split}
	\end{equation}
We can compute the expected values of the trace of powers of $M$ since we can write the trace as products of matrix entries $M_{i,j}$
	\begin{equation}
	\text{Tr} M^n = \sum_{i=1}^n \sum_{(i_1, \dots, i_{k-1}) \in [n]^{k-1}} M_{i, i_1} M_{i_1, i_2} \cdots M_{i_{k-1} , i},
	\end{equation}
and the matrix elements are independent and identically distributed normal random variables. In particular, one can compute the first couple of elements in the expansion explicitly without any trouble:
	\begin{equation}
	\begin{split}
	\mathbb{E}\left[ \text{Tr} M^0\right] &= n\\
	\mathbb{E}\left[ \text{Tr} M^1\right] &= \sum_{i=1}^n \mathbb{E}\left[ M_{i,i}\right] =0 \\
	\mathbb{E}\left[ \text{Tr} M^2\right] &= \sum_{i=1}^n \sum_{j=1}^n\mathbb{E}\left[ | M_{i,j}|^2\right] = n^2\\
	 \mathbb{E}\left[ \text{Tr} M^3\right] &= 0.
	\end{split}
	\end{equation}
In this computation, note that the odd moment are always zero since the odd moments of the the any normal distribution are always zero. In particular,  non-trivial terms arise from terms $M_{i,j}^s M_{j,i}^t$ with $s+t$ even. Consider the case of $j=4$, we have the decomposition into three types of terms
	\begin{equation}
	\begin{split}
	&\mathbb{E}\left[ \text{Tr} M^4\right] =\\
	& \sum_{i=1}^n \left( \sum_{\substack{(i_1, i_2, i_3) \\ i_1 \neq i_3} } + \sum_{\substack{(i_1, i_2, i_3) \\ i_1 = i_3 , i \neq i_2} }   + \sum_{\substack{(i_1, i_2, i_3) \\ i_1 = i_3 , i = i_2} }\right) \mathbb{E} \left[ M_{i,i_1} M_{i_1,i_2} M_{i_2,i_3} M_{i_3,i} \right].
	\end{split}
	\end{equation}
The first sum (inside the sum) is exactly zero since it is a sum of expected values of products of independent random variables with mean zero, but the the terms inside the other two sums are one and three, respectively. That is,
	\begin{equation}\label{eq:tr4}
	\begin{split}
	\mathbb{E}\left[ \text{Tr} M^4\right] &= |\{ (i, i_1, i_2 , i_3) \in [n]^4 | i_1 =i_3 , i \neq i_2\} | \mathbb{E}\left( |X|^2 |Y|^2 \right)\\
	&+3  |\{ (i, i_1, i_2 , i_3) \in [n]^4 | i_1 =i_3 , i = i_2\} | \mathbb{E}(|X|^4)\\
	&= O(n^3) + O(n^2),
	\end{split}
	\end{equation}
with $X$ and $Y$ independent random variable with $\mathcal{N}_{\mathbb{C}}(0,1)$-normal distribution. The sets in the eqn.~\eqref{eq:tr4} have a combinatorial interpretation: elements from the first set are in bijections with all connected graphs on the vertex set $\{1, 2, \dots, n\}$ with two edges and no cycles; and the elements of the second set are in bijection with all connected graphs on the vertex set $\{1, 2, \dots, n\}$ with two edges and exactly one cycle. For general $2j \in 2 \mathbb{N}$, one can write $\mathbb{E}\left[ \text{Tr} M^{2j }\right]$ as sums of different type of terms that are in bijection with connected graphs on the vertex set $\{1, 2, \dots, n\}$ with $j$ edges and certain type of cycles (the full bijection is left as an exercise for the reader, see \cite{Eynard04}). Moreover, for $n\times n$ Hermitian matrices in the GUE, we have that the expected value $\mathbb{E}[\text{Tr }M^{2j}]$  is a polynomial of on the variable $n$ of degree $j+1$. Summarizing, we write 
	\begin{equation}
	\begin{split}
	W_1(z)&=\mathbb{E}\left[ \text{Tr} \frac{1}{z-M/\sqrt{n}} \right] \\
	&= \sum_{g=0}^{\infty} n^{1/2 -g} W_{g,1}(z)\\
	&= \sum_{g=0}^{\infty} \hbar^{2g-1} W_{g,1}(z),
	\end{split}
	\end{equation}
where we have normalized the position of the eigenvalues by dividing $M$ by $\sqrt{n}$ (as in the Wigner semicircle law \ref{wigner}), and we set $\hbar = n^{-1/2}$ in the last line. This story generalizes for the $k$-point function, and one can show
	\begin{equation}
	\begin{split}
	W_{k}(z_1, \dots , z_k) &= \mathbb{E} \left[ \text{Tr}\frac{1}{z_1 - M /\sqrt{n}} \cdots \text{Tr}\frac{1}{z_n - M /\sqrt{n}} \right]_c\\
	&= \sum_{g=0}^\infty \hbar^{2g-2+k} W_{g,k}(z_1, \dots , z_n ).
	\end{split}
	\end{equation}
The \emph{EOC topological recursion} computes the $W_{k,g}$ terms.

	\begin{theorem}[Theorem 4.3 \cite{Motohico}]\label{th:top}
	The $k$-point functions (in the limit $n \rightarrow \infty$) satisfy the EOC topological recursion
	\begin{equation}\label{eq:moto}
	\begin{split}
	&W_{g,k}(t_1, \dots, t_k) =\\
	& = \sum_{a=0,\infty} \underset{x=a}{\text{Res}} K(t_1, t) \left[ W_{n+1, g}(t, -t, t_2, \dots, t_k) + \sum^{\prime}_{\substack{g_1 +g_2 =g \\ I \sqcup J = \{ 2, \dots, n \}} }W_{g_1, |I|+1}(t, t_I)W_{g_1, |J|+1}(-t, t_J) \right] dt_1 \cdots dt_n
	\end{split}
	\end{equation}
where the residue is given computed with respect to the $t_1$ variable, the sum $\sum^{\prime}$ excludes the terms $I = \emptyset, g_1 =0$ and $I = \{z_2, \dots , z_n \} , g_1 = g$, and the kernel is given by
	\begin{equation}
	K(t_1, t) = - \frac{1}{64} \left(\frac{1}{t +t_1} + \frac{1}{t-t_1} \right) \frac{(t^2 -1)^3}{t^2}.
	\end{equation}
	\end{theorem}

\begin{rem}
In Theorem~\ref{th:top}, there is a non-trivial change of coordinates
	\begin{equation}
	W_{g,k}(z_1, \dots, z_k)dz_1 \cdots dz_k \rightarrow  W_{g,k}(t_1, \dots, t_k)dt_1 \cdots dt_k 
	\end{equation}
with $t = \frac{y-1}{y+1}$ and $y$ a solution of the \textit{spectral curve} $y^2 + z y + 1 = 0$, which is obtained as the Stiltjes transform of the Wigner semicircle distribution in this example. Also, note that residue points $a = 0, \infty$ in eqn.~\eqref{eq:moto} are given by the ramification points of the projection of the points from the spectral curve to the complex plane (i.e.~$(z,y)  \mapsto z$). In particular, one should note that all that is needed to compute(recursively) all of the $W_{g,k}$ terms is the first two terms, $W_{0,1}$ and $W_{0,2}$ and the spectral curve (e.g.~$y^2 + z y + 1 = 0$) with a projection map (e.g.~$(z, y) \mapsto z$). 
\end{rem}
In general, the EOC topological recursion has found many other applications outside the scope of matrix model, including enumerative geometry, isomonodromic systems, quantum cohomology (see \cite{Eynard11, Iwaki, Shadrin}). We define it in its original setting of matrix models \cite{Eynard04} and one can find more general definitions in \cite{Bouchard, Motohico14}. Given a matrix model
	\begin{equation}
	Z_{V} = \int_{H_n} dM e^{- \text{Tr} V(M)}
	\end{equation}
where $V: \mathbb{C}[x] \rightarrow \mathbb{C}[x]$ is an even degree polynomial bounded below (i.e.~the coefficient of the highest order term is positive). The topological recursion computes the asymptotic expansion of the $k$-point correlation functions 
	\begin{equation}
	\begin{split}
W_n (z_1, \dots, z_n) &=  \mathbb{E}\left[ \text{Tr}\frac{1}{z_1 - M /\sqrt{n}} \cdots \text{Tr}\frac{1}{z_n - M /\sqrt{n}} \right]_c\\
&= \sum_{g=0}^\infty \hbar^{2g-2+n} W_{g,n}(z_1, \dots , z_n )
	\end{split}
	\end{equation}
In fact, the terms $W_{g,k}$ are computed recursively on other terms with higher the Euler characteristic, $\chi = 2-2g -n$, for $g \geq 0$ and $n >0$. The recursion begins with the terms for $\chi = 1, 0$ (i.e.~$W_{0,1}(z_1)$ and $W_{0,2}(z_1, z_2)$). Moreover, one needs two projection maps
	\begin{equation}
	x,y : \mathbb{C} \rightarrow \mathbb{C},
	\end{equation}
which satisfy an algebraic equation called the \textit{spectral curve} which is given by $y = W_{0,1}(z)$ The recursion kernel is defined by
	\begin{equation}
	K(z_1, z_2) = - \frac{1}{2} \frac{\int_{\bar{z_2}}^{z_2} W_{0,2}(z_1 , z) dz}{W_{0,1}(z_2) - W_{0,1}(\bar{z_2}) },
	\end{equation}
and then one can show that the rest of the $W_{g,k}$'s satisfy the \textit{EOC topological recursion}:
	\begin{equation}
	\begin{split}
	&W_{g,n}(z_1, \dots, z_n) =\\
	&\sum_a \underset{x=a}{\text{Res}} K(z_1, z) \left[ W_{n+1, g}(z, \bar{z}, z_2, \dots, z_n) + \sum^{\prime}_{\substack{g_1 +g_2 =g \\ I \sqcup J = \{ 2, \dots, n \}} }W_{g_1, |I|+1}(z, z_I)W_{g_1, |J|+1}(\bar{z}, z_J) \right] dz_1 \cdots dz_n
	\end{split}
	\end{equation}
with the sum is over the (simple) branch points of $x: \mathbb{C} \rightarrow \mathbb{C}$, $\bar{z}$ a local involution around each branch point $a$, and the sum $\sum^{\prime}$ excluding the terms $\left(I = \emptyset, g_1 =0 \right)$ and $\left( I = \{z_2, \dots , z_n \} , g_1 = g\right)$.

\begin{rem}
This example with $y^2 + xy +1 =0$ was thoroughly studied in \cite{Motohico}.In \cite{Motohico}, the starting point was not matrix models, but rather a generalization of the Catalan numbers. The functions $W_{g,n}(z_1, \dots, z_n)$ in this case are generating functions for the 1-skeletons (up to homotopy) of a genus $g$ Riemann surface with $n$ vertexes. In fact, the case for $W_{(0,1)}(z_1)$ gives you the generating function for the ordinary Catalan numbers, and from a results on Catalan numbers, one can determine the spectral curve $y^2 + xy +1 =0$ explicitly.
\end{rem}

\section*{Acknowledgments}

	AS is extremely thankful for the hospitality and support from the organizers of the \emph{2018 Arizona School of Analysis and Mathematical Physics}: Houssam Abdul-Rahman, Robert Sims, and Amanda Young. AS is also grateful to Eric Brattain, Norman Do, Danilo Lewanski, Motohico Mulase, Paul Norbury, Leonid Petrov, David Renfrew, and Craig Tracy for helpful discussions. AS was partially supported by the NSF grant DMS-1664617.

\bibliographystyle{alpha}

\end{document}